\documentclass[aps,prd,twocolumn,showpacs,nofootinbib,floatfix,superscriptaddress]{revtex4-1}

\newcommand{\Rmnum}[1]{\expandafter\@slowromancap\romannumeral #1@}
\makeatother
\usepackage[colorlinks,linkcolor=blue,anchorcolor=blue,citecolor=blue,urlcolor=blue,breaklinks=true]{hyperref}

\usepackage{epsfig}
\usepackage{natbib}
\usepackage{epstopdf}
\usepackage{mathrsfs}
\usepackage{slashed}
\usepackage{amsmath}
\usepackage{verbatim}
\usepackage{graphicx}
\usepackage{amssymb}
\usepackage{psfrag}
\usepackage{array}
\usepackage{subfig}
\newcommand{\ud}{\mathrm{d}}
\newcommand{\ue}{\mathrm{e}}
\newcommand{\ui}{\mathrm{i}}

\newcommand{\bef}{\begin{figure}[htb]}
\newcommand{\eef}{\end{figure}}

\def\bea#1\eea{\begin{align}#1\end{align}}
\newcommand{\nnu}{\nonumber\\}

\begin{document}
\title{Revisiting heavy quark radiative energy loss in nuclei within the high-twist approach}

\author{Yi-Lun Du}
\affiliation{Department of Physics, Nanjing University, Nanjing 210093, China}
\affiliation{Frankfurt Institute for Advanced Studies, Giersch Science Center, D-60438 Frankfurt am Main, Germany}
\affiliation{Institut f$\ddot{u}$r Theoretische Physik, Goethe Universit$\ddot{a}$t Frankfurt, D-60438 Frankfurt am Main, Germany}

\author{Yayun He}
\affiliation{Key Laboratory of Quark and Lepton Physics (MOE) and Institute of Particle Physics, Central China Normal University, Wuhan 430079, China}
\affiliation{Nuclear Science Division, Lawrence Berkeley National Laboratory, Berkeley, California 94720, USA}

\author{Xin-Nian Wang}
\affiliation{Key Laboratory of Quark and Lepton Physics (MOE) and Institute of Particle Physics, Central China Normal University, Wuhan 430079, China}
\affiliation{Nuclear Science Division, Lawrence Berkeley National Laboratory, Berkeley, California 94720, USA}

\author{Hongxi Xing}
\affiliation{School of Physics and Telecommunication Engineering, South China Normal University, Guangzhou 510006, China}
\affiliation{Department of Physics and Astronomy, Northwestern University,  Evanston, Illinois 60208, USA}
\affiliation{High Energy Physics Division, Argonne National Laboratory, Argonne, Illinois 60439, USA}

\author{Hong-Shi Zong}
\affiliation{Department of Physics, Nanjing University, Nanjing 210093, China}
\affiliation{Joint Center for Particle, Nuclear Physics and Cosmology, Nanjing 210093, China}
\affiliation{State Key Laboratory of Theoretical Physics, Institute of Theoretical Physics, CAS, Beijing 100190, China}

\begin{abstract}
We revisit the calculation of multiple parton scattering of a heavy quark in nuclei within the framework of recently improved high-twist factorization formalism, in which gauge invariance is ensured by a delicate setup of the initial partons' transverse momenta. We derive a new result for medium modified heavy quark fragmentation functions in deeply inelastic scattering. It is consistent with the previous calculation of light quark energy loss in the massless limit, but leads to a new correction term in the heavy quark case, which vanishes in the soft gluon radiation limit.  We show numerically the significance of the new correction term in the calculation of heavy quark energy loss as compared to previous studies and with soft gluon radiation approximation.
\end{abstract}

\maketitle

\section{Introduction}
\label{sec:intro} 
In the past decade, tremendous progress has been made in understanding the jet quenching phenomena via various observables in high-energy nucleus-nucleus collisions \cite{Gyulassy:2003mc,Majumder:2010qh,Qin:2015srf}. Within perturbative-QCD-based theoretical frameworks for multiple parton scattering and parton energy loss in a nuclear medium, one is able to use the experimental data on jet quenching to probe the fundamental properties of the cold nuclei and the hot dense medium created in high-energy heavy-ion collisions \cite{Wang:2002ri,Bass:2008rv,Armesto:2009zi,Chen:2010te}. One seminal study is the  systematic extraction of jet quenching parameter, $\hat q$, by the JET Collaboration \cite{Burke:2013yra}, in which a global fitting to the experimental data on single inclusive hadron production at the Relativistic Heavy-Ion Collider and  the Large Hadron Collider has been performed based on several jet quenching models for multiple parton scattering and parton energy loss. 

Several of the successful jet quenching models in explaining the experimental data on hadron or jet production are based on the high-twist expansion approach \cite{Guo:2000nz,Wang:2001ifa,Zhang:2003yn,Majumder:2009ge}, where contributions from multiple scattering between a propagating jet and medium partons can be effectively factorized as higher twist corrections to the vacuum fragmentation functions. These models utilize the generalized twist-4 factorization formalism developed by Qiu and Sterman \cite{Luo:1992fz, Luo:1994np}. The first calculation within this framework is performed in the process of semi-inclusive electron-nucleus deep inelastic scattering (SIDIS), where the light quark energy loss is encoded into the medium modified quark fragmentation functions \cite{Guo:2000nz, Wang:2001ifa}. This calculation has been further improved by going beyond the helicity amplitude approximation \cite{Zhang:2003yn} and including multiple scattering \cite{Majumder:2009ge} and double quark scattering \cite{Schafer:2007xh}. The same techniques were also applied to evaluate heavy quark energy loss in SIDIS by considering charged-current interaction \cite{Zhang:2003wk, Zhang:2004qm}. Similarly, one can also study the effect of initial state parton energy loss within the same framework, which has been evaluated in Drell-Yan dilepton production in proton-nucleus collisions. In analogy to the calculation in SIDIS, the effect of initial state parton energy loss in the Drell-Yan process is encoded in the medium modified beam quark distribution function \cite{Xing:2011fb}. In all these calculations, the main goal is to investigate the effect of parton energy loss in nuclear medium from either final state or initial state multiple scatterings. Therefore, only a subset of Feynman diagrams at next-to-leading order (NLO) are considered to simplify the calculation by choosing appropriate physical gauge for the radiated gluon. However, without the inclusion of the complete NLO Feynman diagrams, a consistent check of gauge invariance is impossible. 

The first complete NLO calculation at twist 4 has been carried out for the transverse momentum weighted differential cross section in SIDIS \cite{Kang:2013raa, Kang:2014ela}, and later extended to Drell-Yan lepton pair production in proton-nucleus collisions \cite{Kang:2016ron}. The calculated observable is directly related to the transverse momentum broadening in SIDIS off nuclear targets and heavy-ion collisions. In these two calculations, the authors have included all Feynman diagrams whose contributions are enhanced by the nuclear size. One technical aspect of these twist-4 NLO calculations is that appropriate initial partons' transverse momenta flow has to be assigned to ensure the gauge invariance during the course of collinear expansion. This is of particular importance for the subprocess of interference between soft and hard rescatterings. The calculation of symmetric subprocesses (soft-soft and hard-hard double scattering) is less ambiguous; one can use either way to reach a gauge invariant result. 

In this paper, we apply the newly developed twist-expansion technique to recalculate the effect of final state parton energy loss in SIDIS. In particular, we focus on the channel of charged-current interaction. This allows us to study light quark and heavy quark radiative energy loss on the same footing. We compare our results with the previous studies for both the light quark and heavy quark.  

The rest of this paper is organized as follows. In Sec.~\ref{sec-framework}, we introduce our notations, and review the generalized factorization formalism at twist 4. In Sec.~\ref{sec-calculation}, we present the details of our calculation at twist 4 by including both the quark-gluon double scattering and the interference between single and triple scatterings. We show how the principle of gauge invariance guides the transverse momentum flow of the four initial partons from the nuclear target. In Sec.~\ref{sec-MFF}, we give the final result of medium modified heavy quark fragmentation functions. In Sec.~\ref{sec-energy loss}, we illustrate numerically the importance of our gauge invariant result for heavy quark energy loss by comparing with the previous study and in the soft limit. A summary is given in Sec.~\ref{sec-sum}.

\section{General framework}
\label{sec-framework}


For simplicity, we consider the following process of heavy quark production via the charged-current interaction in deep inelastic scattering (DIS) off a large nucleus $A$,
\bea
L(\ell_1) + A(p)\to \nu_L(\ell_2) + H(\ell_H) + X,
\eea
where $\ell_1$ and $\ell_2$ are the momenta of the incoming lepton and the outgoing neutrino, $p=[p^+,0,\bold{0}_\perp]$ is the momentum per nucleon of the target nucleus with atomic number $A$, and $\ell_H$ is the observed final state heavy meson (H) momentum. In the channel of charged-current interaction, the momentum transfer via the exchange of a $W^{\pm}$ boson is given by $q=\ell_1-\ell_2$ with the invariant mass $q^2=-Q^2$ and $Q^2\ll M_W^2$ is assumed. Notice that both the heavy quark flavor and the momentum scale of the exchanged vector boson are labeled as $Q$ in this work. 

The differential cross section for single inclusive heavy meson production can be written as 
\bea
E_{2}E_{H}\frac{{\ud}\sigma^H}{d^3 \ell_2 d^3 \ell_H}=\frac{\mathrm{G_F^2}}{(4\pi)^3 s}L_{\mu\nu} E_{H}\frac{dW^{\mu\nu}}{d^3\ell_H},
\eea
where $s = (p+\ell_1)^2$ is the lepton-nucleon collision energy and $\mathrm{G_F}$ stands for the four-fermion coupling constant. The charged-current leptonic tensor reads 
\bea
L_{\mu\nu} =\frac{1}{2}\mathrm{Tr}\left[\slashed{\ell}_1\gamma_\mu(1-\gamma_5)\slashed{\ell}_2(1+\gamma_5)\gamma_\nu\right].
\label{eq:L}
\eea
The semi-inclusive hadronic tensor is defined by
\bea
\begin{aligned}
E_{H}\frac{{\ud}W^{\mu\nu}}{{\ud}^3\ell_H}=&\frac{1}{2}\sum_X\langle A\mid J^\mu\mid X,H\rangle\langle X,H\mid J^{\nu\dagger} \mid A\rangle\\
&\times2\pi\delta^4(q+p-p_X-\ell_H),
\end{aligned}
\eea
where $\sum_X$ sums over all possible final states, $J^\mu=\sum_f \bar{\psi}_f\gamma^\mu V\psi_f$ is the hadronic charged current, and $V = (1-\gamma_5)V_{ij}$ with $V_{ij}$ stands for the Cabibbo-Kobayashi-Maskawa flavor mixing matrix~\cite{Aivazis:1993kh}. 

\begin{figure*}[t] 
\centering
\subfloat[][\empty]{
\begin{minipage}[t]{0.4\linewidth} 
\centering 
\includegraphics[width=2.8in]{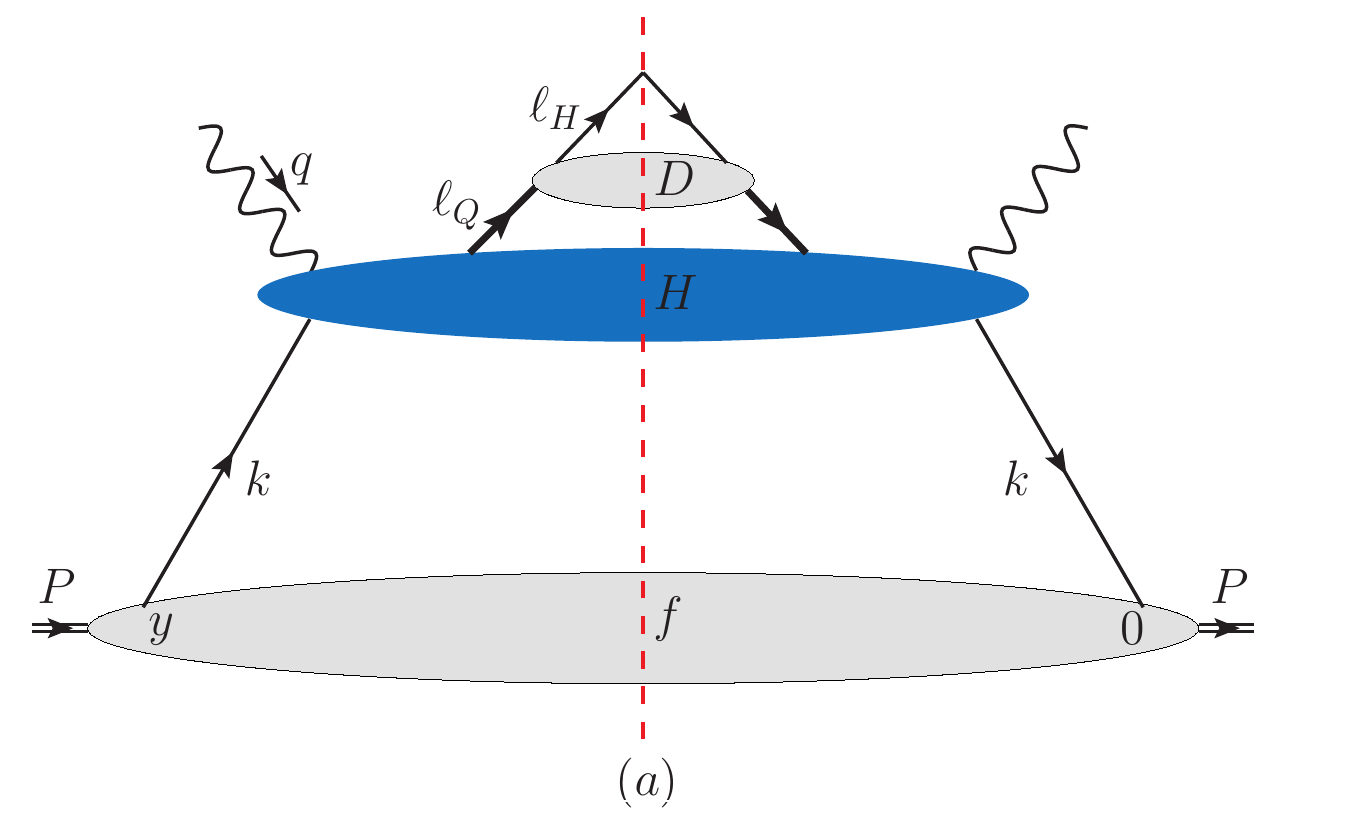} 
\end{minipage}\label{fig-DIS.a}
} 
\subfloat[][\empty]{
\begin{minipage}[t]{0.4\linewidth} 
\centering 
\includegraphics[width=2.8in]{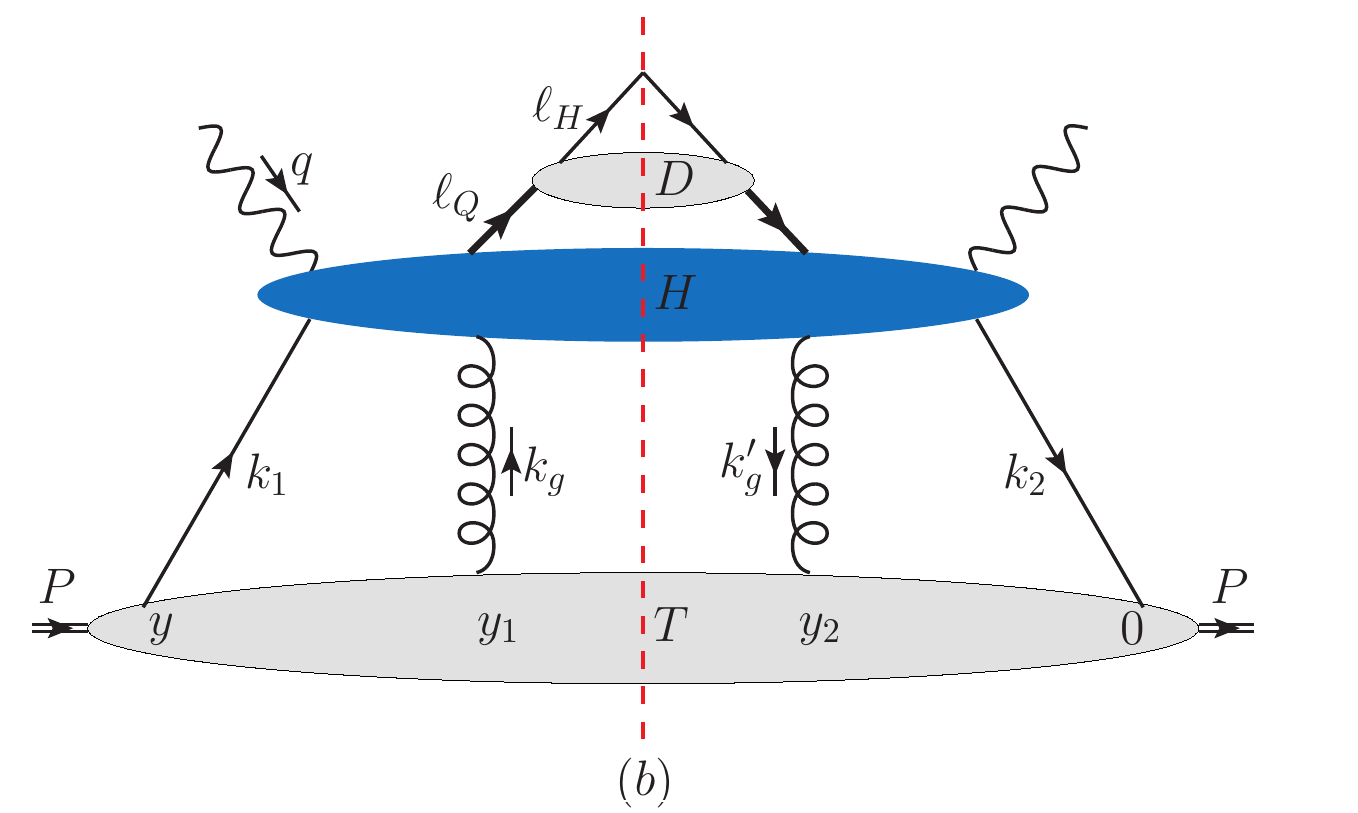} 
\end{minipage}\label{fig-DIS.b} 
}
\caption{ The general diagrams for single inclusive heavy meson production in SIDIS off a nuclear target: (a) single scattering and (b) heavy-quark-gluon double scattering. The thick lines represent heavy quarks.}
\label{fig-DIS}
\end{figure*}

The above mechanism of single inclusive heavy meson production in DIS can be illustrated in Fig.~\ref{fig-DIS.a}: a $W^\pm$ boson is radiated from the projectile lepton and collides with a light quark from the nuclear target, a heavy quark $Q$ with mass $M$ and momentum $\ell_Q$ is produced and then fragments into a heavy meson in the final state. In fixed order calculation, there is no collinear divergence for final state gluon radiation, because of the finite heavy quark mass that naturally serves as a regulator. However, in the case when the momentum scale $Q$ is much larger than the heavy quark mass $M$, $Q^2\gg M^2$, one encounters large logarithms, $\ln(Q^2/M^2)$, which spoil the perturbative convergence. In this case, an all order resummation of such large logarithms has to be performed in the calculation. Such a resummation is normally done by solving the renormalization equation of final state heavy quark fragmentation functions \cite{Mele:1990cw}. 
Therefore, the hadronic tensor at leading twist can be factorized into the convolution of nuclear quark distributions $f_q^A$, heavy flavor fragmentation functions $D_{Q\to H}$, and the partonic cross section $H^{\mu\nu}_{W^{\pm}+q\to Q+X}$,
\begin{widetext}
\bea
\frac{dW^{\mu\nu,S}}{dz_H} =  f_q^A(x,\mu_I^2)\otimes H^{\mu\nu}_{W^{\pm}+q\to Q+X}(x,z_H,p,q,M,\mu_I,\mu_F)\otimes D_{Q\to H}(z_H,\mu_F^2),
\eea
where the superscript $S$ stands for single scattering, the sum over initial state quark flavors is suppressed for simplicity, and $\mu_I$ and $\mu_F$ represent initial and final state factorization scale, respectively.

At leading twist, the hard partonic part at lowest order is
\bea
\label{H0}
H_{\mu\nu}^{(0)}(x,p,q,M)=&\frac{1}{2}|V_{ij}|^2\mathrm{Tr}\left[\slashed{p}\gamma_\mu (1-\gamma_5)(\slashed{q}+x\slashed{p}+M)(1+\gamma_5)\gamma_\nu\right] \frac{2\pi}{2p\cdot q}\delta(x-x_B-x_M),
\eea
where the Lorentz invariant variables are defined as
\bea
x_M = \frac{M^2}{2p\cdot q}, ~~~ x_B = \frac{Q^2}{2p\cdot q}, ~~~z = \frac{p\cdot \ell_Q}{p\cdot q}, ~~~z_H = \frac{p\cdot \ell_H}{p\cdot q}.
\label{eq-xMxB}
\eea
\end{widetext}
 
Inside a large nucleus, the propagating heavy quark encounters additional scatterings with the nuclear target remnants as shown in Fig.~\ref{fig-DIS.b}, which lead to nontrivial medium modifications to heavy quark production. In this paper, we focus on the radiative energy loss due to the medium induced gluon radiations with 4-momenta $\ell$. This effect in general is a nuclear enhanced power correction to final state heavy quark fragmentation functions \cite{Zhang:2003wk, Zhang:2004qm}. Such a power correction can be computed within the high-twist expansion formalism developed by Qiu and Sterman \cite{Luo:1992fz, Luo:1994np}. Within this framework, the collinear QCD dynamics of multiple parton interaction are contained in the medium modified splitting functions that are perturbatively calculable, while the medium property is contained in the high-twist nonperturbative multiparton correlation functions. Recently, the QCD evolution equation for the renormalized twist-4 quark-gluon correlation function was derived ~\cite{Kang:2013raa}. In this paper, we aim to derive the perturbative medium modified splitting functions through NLO computations.  

We apply the recently improved twist-4 collinear factorization technique ~\cite{Kang:2014ela} to calculate the medium induced gluon radiation of the heavy quark in DIS. The leading contribution from double scattering processes can be obtained by taking collinear expansion of the hard partonic cross section with respect to the transverse momenta of initial partons
\begin{widetext}
\bea
\label{eqn:tensorw}
\frac{{\ud}W_{\mu\nu}^D}{{\ud}z_H}=&\sum_q \int_{z_H}^1\frac{{\ud}z}{z}D_{Q\to H}(z_H/z)\int\frac{{\ud}y^-}{2\pi}{\ud}y_1^-{\ud}y_2^-\frac{1}{2}\langle A\mid\bar{\psi}_q(0)\gamma^+{F_\sigma}^+(y_2^-)F^{+\sigma}(y_1^-)\psi_q(y^-)\mid A\rangle\nnu
&\times\left(-\frac{1}{2}g^{\alpha\beta}\right)\left[\left.\dfrac{\partial^2}{\partial k_{2T}^{\alpha}\partial k_{3T}^{\beta}}\bar{H}^D_{\mu\nu}(y^-,y_1^-,y_2^-,k_{2T},k_{3T}, p, q, M, z)\right]\right |_{\substack{k_{2T}=0\\k_{3T}=0}},
\eea
where the superscript $D$ stands for double scattering, and $k_{2T}$ and $k_{3T}$ are the relative transverse momenta carried by gluons from the nucleus in the double scattering (see the next section for their definitions) . The hard partonic part of central-cut diagrams can be written in the following general form:
\bea
\bar{H}^\mathrm{D}_{\mathrm{C}\,\mu\nu}(y^-,y_1^-,y_2^-,k_{2T},k_{3T}, p, q, M, z)=&\int{\ud}x\frac{{\ud}x_1}{2\pi}\frac{{\ud}x_2}{2\pi}{\ue}^{{\ui}x_1p^+y^-+{\ui}x_2p^+y_1^-+{\ui}(x-x_1-x_2)p^+y_2^-}\int\frac{{\ud}^4\ell}{(2\pi)^4}\nnu
&\times\frac{1}{2}\mathrm{Tr}\left[p\cdot\gamma\gamma_\mu V p^\sigma p^\rho\hat{H}_{\sigma\rho}V^\dagger\gamma_\nu\right]2\pi\delta_+(\ell^2)\delta(1-z-\frac{\ell^-}{q^-}).
\eea
\end{widetext}
We apply collinear approximation to simplify the evaluation of the trace of $\gamma$ matrices, 
\bea
p^\sigma \hat{H}_{\sigma\rho}p^\rho\approx\frac{(\gamma\cdot \ell_Q+M)}{4\ell_Q^-}\mathrm{Tr}\left[\gamma^-p^\sigma\hat{H}_{\sigma\rho}p^\rho\right].
\eea
After integrating out $x, x_1, x_2$, and $\ell^{\pm}$ with the help of contour integration and $\delta$-functions from final state phase space, the partonic hard part at twist-4 can be factorized into the product of leading-order hard part for $V$+quark $H_{\mu\nu}^{(0)}(x, p, q, M)$ at leading twist as shown in Eq.~(\ref{H0}) and the heavy-quark-gluon rescattering part $\bar{H}^\mathrm{D}$,
\begin{widetext}
\bea\label{factorized}
\begin{aligned}
&\bar{H}^D_{\mu\nu}(y^-,y_1^-,y_2^-,k_{2T},k_{3T}, p, q, M, z)= \int{\ud}xH_{\mu\nu}^{(0)}(x,p,q,M)\bar{H}^D(y^-,y_1^-,y_2^-,k_{2T},k_{3T}, p, q, M, z).
\end{aligned}
\eea
\end{widetext}
Hence in the following we only show the rescattering part $\bar{H}^\mathrm{D}$. Notice that we have verified that the same result can be obtained by going beyond the collinear approximation, i.e., through exact calculation by contracting the hadronic tensor $W_{\mu\nu}$ with the so-called ``metric" contribution proportional to $g^{\mu\nu}$ in the charged-current leptonic tensor $L^{\mu\nu}$ in Eq.(\ref{eq:L}) as in Ref.~\cite{Kang:2014ela}. 

\section{Medium induced gluon radiation at twist 4}
\label{sec-calculation}

\begin{figure*} [t]
\centering
\subfloat[][\empty]{
\begin{minipage}[t]{0.3\linewidth} 
\centering 
\includegraphics[width=2.0in]{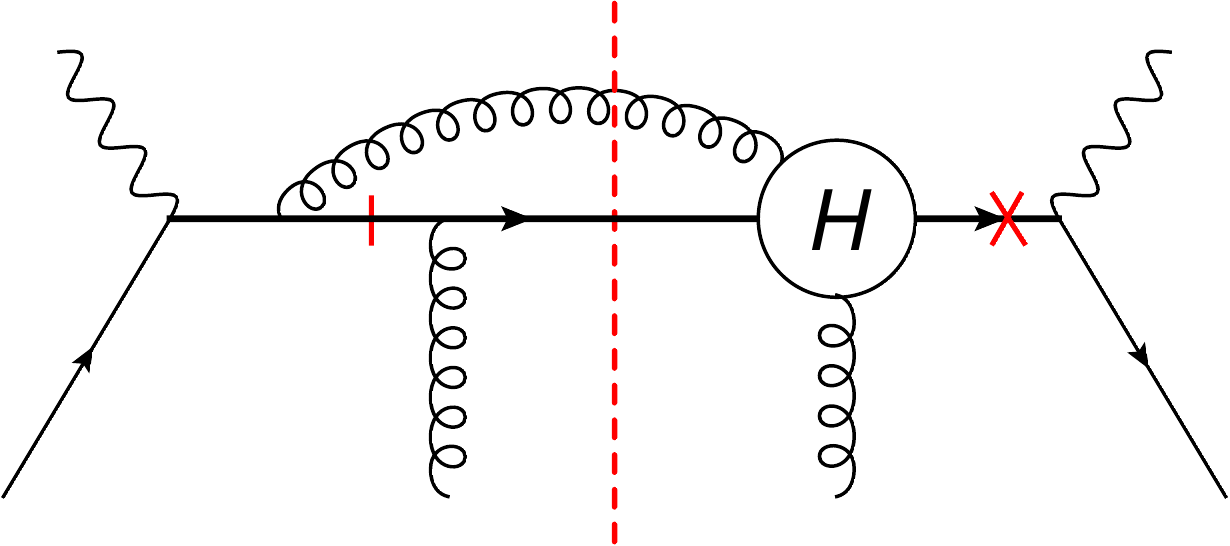} 
\end{minipage}%
} 
\subfloat[][\empty]{
\begin{minipage}[t]{0.7\linewidth} 
\centering 
\includegraphics[width=4.5in]{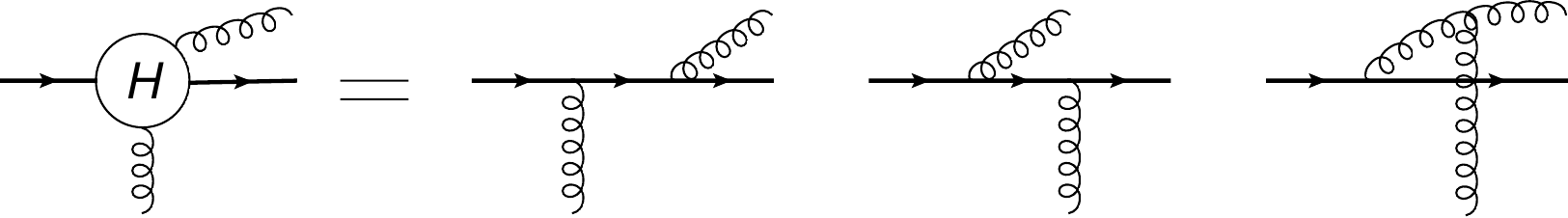} 
\end{minipage} 
}\caption{ Left: The central-cut diagrams for soft-hard double scatterings in SIDIS. The short bar and cross mark indicate the propagator where the soft pole and hard pole arise. Right: explicit diagrams for the ``$H$"-blob representing the $Q+g\to Q+g$ process.}
\label{fig-sh}
\end{figure*} 

In this section we present the details of the calculation of the twist-4 contribution to heavy meson production in DIS. A complete twist-4 calculation at NLO contains contributions involving both quark-gluon and gluon-gluon twist-4 matrix elements. In this paper, we neglect contribution from gluon-gluon double scattering as it is irrelevant to the power correction to final state heavy quark fragmentation, which is the main focus of this paper. There are four different kinds of subprocesses in quark-gluon double scattering with real gluon radiation in the final state: soft-soft double scattering, hard-hard double scattering, and the interferences between them, referred as soft-hard and hard-soft double scattering.\footnote{We follow the terminologies in Ref.~\cite{Kang:2014ela}, which are different from those in Ref.~\cite{Zhang:2004qm}.} This classification is based on whether the momenta of the exchanged gluons (i.e., $k_g^{}$ and $k_g^\prime$ on the amplitude and complex conjugate of the amplitude, respectively) have 0 or finite momenta in the collinear limit $k_{2T}=k_{3T}=0$.

The computation of heavy quark energy loss in DIS has already been performed in Ref.~\cite{Zhang:2004qm} by setting the same transverse momentum for the rescattered gluons from the nucleus. This setting has been shown to be valid, in the case of light hadron production, for symmetric double scattering subprocesses, such as hard-hard and soft-soft double scatterings \cite{Kang:2014ela}. We have verified explicitly that this statement also holds for heavy flavor meson production. Therefore we do not present exhaustive calculations of the subprocesses for hard-hard and soft-soft double scattering, but instead concentrate on the asymmetric subprocesses, i.e., the interference between soft and hard rescatterings. In order to ensure gauge invariant results at twist-4, on-shell conditions for the initial partons (associated with the $2\to 2$ hard scattering) have to be satisfied up to $\mathcal O(k_{2T})$ and $\mathcal O(k_{3T})$ as specified in Ref.~\cite{Kang:2014ela} for light hadron production, which is also true for heavy meson production as explained below. This requirement is fulfilled, as proposed in Ref.~\cite{Kang:2014ela}, by a delicate assignment of the transverse momenta for the initial partons from the nuclear target. We apply the same assignment in the calculation of interference subprocesses for heavy meson production in DIS.

To simplify the calculation by means of less Feynman diagram evaluation, we use the following polarization tensor for the final state radiated gluon in light-cone gauge ($n\cdot A=0$),
\bea
\quad \varepsilon_{\mu\nu}(k)=-g_{\mu\nu} + \frac{k_\mu n_\nu+n_\mu k_\nu}{k\cdot n},
\label{eq-gauge}
\eea
where a particular reference vector $n=[1,0^-,\vec{0}_{\perp}]$ has to be implemented. In principle, the final result is gauge independent if one considers all Feynman diagrams. In this paper, we aim to extract the large logarithmic term, $\ln(Q^2/M^2)$, which is only related to final state gluon radiation. Thus we only consider Feynman diagrams with gluon radiated from the final state heavy quark. This simplification is realized by choosing an appropriate light-cone reference vector as specified above. With this particular gauge choice, all Feynman diagrams with gluon radiated from the initial state light quark, as well as the interference between initial and final state gluon radiations, contribute to collinear divergence (in $q_T^2 \ll Q^2$ limit) related to the renormalization of initial state multiparton correlation function, which is not the focus of this paper.   

\subsection{Heavy-quark-gluon double scattering}

With the light-cone gauge choice for the radiated gluon as shown in Eq.~(\ref{eq-gauge}), half of the diagrams that we need to consider in the process of interference between soft and hard gluon rescattering are shown in Fig.~\ref{fig-sh}.  All the relevant double scattering diagrams can be found in Refs.~\cite{Wang:2001ifa,Zhang:2004qm}. The soft rescattering subprocess at amplitude level can be viewed as two factorized scatterings: the first $V+q\to Q+g$ hard scattering, and the second soft rescattering $Q+g\to Q$ by exchanging a soft gluon. To ensure gauge invariance, the initial quark associated with the first scattering is required to be on its mass shell, i.e., $k_1^2 = 0$~\cite{Peskin:1995ev}, while no on-shell requirement is needed for the initial gluon in the second soft scattering since it is essentially a leading-order QCD interaction. On the other hand, the hard rescattering subprocess can be factorized as the first vector boson-quark hard scattering $V+q\to Q$ and the second hard rescattering $Q+g\to Q+g$. In this case, to ensure gauge invariance for the second $Q+g\to Q+g$ process, the exchanged gluon is required to be on its mass shell ($k_g^2=0$) up to the order in which we perform collinear expansion, and no requirement is needed for the initial quark associated with the first scattering. In addition, to avoid complications from the ``minus" components of the momenta for both exchanged gluons in the amplitude and its complex conjugate, we could simply set the momenta of the two exchanged gluons as $k_g=x_3p + k_{3T}$ and $k_g^\prime=x_2p + k_{2T}$. To preserve momentum conservation on both sides of the cut line, we have to assign a transverse momentum of $k_{3T}-k_{2T}$ to the initial quark associated with the leading-order $V+q\to Q$ hard scattering. 

We take soft-hard double scattering as an example to outline the essential steps in the calculation of twist-4 contributions, and all the other subprocesses could be evaluated in the same manner. In this subprocess as shown in Fig.~\ref{fig-sh}, based on the analysis specified above, we can assign the momenta for the initial partons as $k_1=xp, k_2=x_1p+k_{3T}-k_{2T} , k_g = x_3p+k_{3T}$ and $k_g^\prime=x_2p+k_{2T}$ [see Fig.~\ref{fig-DIS.b}]. The two on-shell propagators labeled by the short bar and cross mark that lead to soft and hard initial gluons can be expressed as follows:
\begin{widetext}
\bea
\frac{1}{(\ell_Q -k_g)^2-M^2+ i\epsilon} =& \frac{1}{2p^+q^-z(x-x_L-x_B-x_M/z)+ i\epsilon},  \\
\frac{1}{(k_2+q)^2-M^2 - i\epsilon} =& \frac{1}{2p^+q^-(x_1-x_B-x_F-x_M)- i\epsilon}.
\eea
On the other hand, the on-shell condition for the final state quark gives
\bea
\delta_+(\ell_Q^2-M^2) = \frac{1}{2p^+q^-z}\delta(x_1+x_2-x_L-x_D-x_B-x_M/z),
\eea
where $x_M$ and $x_B$ are defined in Eq.~(\ref{eq-xMxB}), and 
\bea
x_L=\frac{\vec{\ell}_T^2}{2p^+q^-z(1-z)},~~~x_D=\frac{\vec{k}_{3T}^2-2\vec{k}_{3T}\cdot\vec{\ell}_T}{2p^+q^-z},~~~x_F=\frac{(\vec{k}_{3T}-\vec{k}_{2T})^2}{2p^+q^-}.
\eea
One can then integrate over $x,~x_1$, and $x_2$,
\bea
&\int {\ud}x\frac{{\ud}x_1}{2\pi}\frac{{\ud}x_2}{2\pi} {\ue}^{{\ui}x_1p^+y^-+{\ui}x_2p^+y_1^-+{\ui}(x-x_1-x_2)p^+y_2^-}\frac{1}{x-x_L-x_B-x_M/z+ i\epsilon}\frac{\delta(x_1+x_2-x_L-x_D-x_B-x_M/z)}{x_1-x_B-x_F-x_M- i\epsilon}\nnu
&={\ue}^{{\ui}(x_B+x_L+x_M/z)p^+y^-}{\ue}^{{\ui}x_Dp^+(y_1^--y_2^-)}\theta(-y_2^-)\theta(y^--y_1^-){\ue}^{-{\ui}(x_L+(1-z)x_M/z-x_F)p^+(y^--y_1^-)},
\label{eq-intg}
\eea
where two of the integrations are carried out by contour integrations, and the last one is fixed by the $\delta$ function from final state on-shell condition. Then the initial state partons' momentum fractions are fixed as follows, 
\bea
x=x_L+x_B+x_M/z,~~~ x_1 = x_B+x_F+x_M,~~~ x_2=x_L+x_D+(1-z)x_M/z-x_F,~~~ x_3=x_D.
\label{eq-fracs}
\eea
\end{widetext}
It is clear that in the collinear limit $k_{2T}=k_{3T}=0$, the momentum fraction for the initial gluon on the left- and right-hand side of the cut line becomes 0 and remains finite, respectively. This explains why we refer to this process as soft-hard double scattering.

The key point in high-twist calculation is to perform collinear expansion. With the fixed momentum fractions for soft-hard double scattering as shown in Eq. (\ref{eq-fracs}), we have
\begin{widetext}
\bea
\left.\dfrac{\partial^2\bar{H}^\mathrm{D}_\mathrm{C-sh}}{\partial k_{2T}^{\alpha}\partial k_{3T}^{\beta}}\right |_{\substack{k_{2T}=0\\k_{3T}=0}}
=&
\int\!{\ud}\vec{\ell}_{T}^2\frac{\alpha_s}{2\pi}\frac{1+z^2}{1-z}\frac{\mathrm{2C_A}\vec{\ell}_T^4}{\left[\vec{\ell}_T^2+(1-z)^2M^2\right]^4}{\ue}^{{\ui}(x+x_L+(1-z)x_M/z)p^+y^-}\frac{2\pi\alpha_s}{N_c}\theta(-y_2^-)\theta(y^--y_1^-) \tilde{H}^\mathrm{D}_\mathrm{C-sh},
\eea
where
\bea
\tilde{H}^\mathrm{D}_\mathrm{C-sh} = &\Bigg\{-1+\frac{(1-z)}{2}+\frac{2(1-z)^3z(1+z)}{1+z^2}\frac{M^2}{\vec{\ell}_T^2}-\frac{(1-z)^4(3z^3-5z^2+7z-1)}{2(1+z^2)}\frac{M^4}{\vec{\ell}_T^4}\nnu
&-\mathrm{\frac{2C_F}{C_A}}\left[(1+z)^2+(1-z)^4\frac{M^2}{\vec{\ell}_T^2}\right]\frac{(1-z)^4}{1+z^2}\frac{M^2}{\vec{\ell}_T^2}\Bigg\}{\ue}^{-{\ui}(x_L+(1-z)x_M/z)p^+(y^--y_1^-)}.
\label{eq-Hcsh}
\eea 
\end{widetext}
Notice that we have neglected contributions that are power suppressed, such as the derivative terms on the twist-4 quark-gluon correlation functions, and that all the momentum fractions ($x$'s) appearing in the phase factor here and below have been shifted to $x_B + x_M$ as Eq.~(\ref{H0}) requires.

Now, let us compare our result to the one in Ref.~\cite{Zhang:2004qm}, which is derived from the naive setup $k_{2T}=k_{3T}$.\footnote{There are some typos in Ref.~\cite{Zhang:2004qm}; the correct final results can be obtained from Eqs.~(\ref{eq-Hcsh})--(\ref{eqn:deri2}) and ~(\ref{eq-l})--(\ref{eq-diff-r}).} The difference from soft-hard double scattering is 
\begin{widetext}
\bea
\Delta \tilde{H}_\mathrm{C-sh}^\mathrm{D} = &\tilde{H}_\mathrm{C-sh}^\mathrm{D} - \tilde{H}_\mathrm{C-sh}^{\mathrm{D}, \text{Ref.~\cite{Zhang:2004qm}}}\nnu
=&-\left[z-\frac{1}{2}+\mathrm{\frac{C_F}{C_A}}(1-z)^2\right]\left[(1+z)^2+(1-z)^4\frac{M^2}{\vec{\ell}_T^2}\right]\frac{(1-z)^2}{1+z^2}\frac{M^2}{\vec{\ell}_T^2}{\ue}^{-{\ui}(x_L+(1-z)x_M/z)p^+(y^--y_1^-)}.
\label{eq-diff-sh}
\eea
\end{widetext}
As one can see the difference is proportional to $(1-z)^2 M^2$, which vanishes in the soft limit $z\to 1$ or massless limit $M=0$.   

Following the same logic, one can assign initial state parton momenta $k_1=xp+k_{2T}-k_{3T}, k_2=x_1p, k_g = x_3p+k_{3T}$ and $k_g^\prime=x_2p+k_{2T}$ in the process of hard-soft double scattering. The final result can be obtained from the soft-hard double scattering via the replacement ${\ue}^{-{\ui}(x_L+(1-z)x_M/z)p^+(y^--y_1^-)}\to{\ue}^{-{\ui}(x_L+(1-z)x_M/z)p^+y_2^-}$ in Eq.~(\ref{eq-Hcsh}). Thus the difference to Ref. \cite{Zhang:2004qm} in hard-soft double scattering is
\begin{widetext}
\bea
\Delta \tilde{H}_\mathrm{C-hs}^\mathrm{D}&=-\left [z-\frac{1}{2}+\mathrm{\frac{C_F}{C_A}}(1-z)^2\right ]\left [(1+z)^2+(1-z)^4\frac{M^2}{\vec{\ell}_T^2}\right ]\frac{(1-z)^2}{1+z^2}\frac{M^2}{\vec{\ell}_T^2}{\ue}^{-{\ui}(x_L+(1-z)x_M/z)p^+y_2^-},
\label{eq-diff-hs}
\eea
\end{widetext}
which, again, vanishes in the soft or massless limit.

For soft-soft and hard-hard double scattering subprocesses, one can use the same assignments of the initial partons' transverse momentum flow as in Ref.~\cite{Zhang:2004qm}, and we obtain the same results, namely, 
\bea
\Delta \tilde{H}_\mathrm{C-ss}^\mathrm{D} = \Delta \tilde{H}_\mathrm{C-hh}^\mathrm{D} = 0.
\label{eq-diff-ss}
\eea
For hard-hard double scattering, we have checked that three different settings of the transverse momenta for initial state partons reach exactly the same result.

By combining all contributions from quark-gluon double scattering with a central cut together, we obtain the final result 
\begin{widetext}
\bea
\left.\dfrac{\partial^2\bar{H}^\mathrm{D}_\mathrm{C}}{\partial k_{2T}^{\alpha}\partial k_{3T}^{\beta}}\right |_{\substack{k_{2T}=0\\k_{3T}=0}}
=&
\int\!{\ud}\vec{\ell}_{T}^2\frac{\alpha_s}{2\pi}\frac{1+z^2}{1-z}\frac{\mathrm{2C_A}\vec{\ell}_T^4}{\left[\vec{\ell}_T^2+(1-z)^2M^2\right]^4}{\ue}^{{\ui}(x+x_L+(1-z)x_M/z)p^+y^-}\frac{2\pi\alpha_s}{N_c}\theta(-y_2^-)\theta(y^--y_1^-) \tilde{H}^\mathrm{D}_\mathrm{C},\label{eqn:ccderi}
\eea
where
\bea\label{eqn:deri2}
\tilde{H}^\mathrm{D}_\mathrm{C}=&\left[1+\frac{(1-z)^2(1-6z+z^2)}{1+z^2}\frac{M^2}{\vec{\ell}_T^2}+\frac{2(1-z)^4z}{1+z^2}\frac{M^4}{\vec{\ell}_T^4}\right]\nnu
&+\!\mathrm{\left[1\!-\!(1-z)\!+\!\frac{C_F}{C_A}(1-z)^2\right]}\left[1+\frac{2(1-z)^4}{1+z^2}\frac{M^2}{\vec{\ell}_T^2}\!+\!(1-z)^4\frac{M^4}{\vec{\ell}_T^4}\right]{\ue}^{-{\ui}(x_L+(1-z)x_M/z)p^+(y^--y_1^-+y_2^-)}\nnu
&+\Bigg\{-1+\frac{(1-z)}{2}+\frac{2(1-z)^3z(1+z)}{1+z^2}\frac{M^2}{\vec{\ell}_T^2}-\frac{(1-z)^4(3z^3-5z^2+7z-1)}{2(1+z^2)}\frac{M^4}{\vec{\ell}_T^4}\nnu
&-\mathrm{\frac{2C_F}{C_A}}\left[(1+z)^2+(1-z)^4\frac{M^2}{\vec{\ell}_T^2}\right]\frac{(1-z)^4}{1+z^2}\frac{M^2}{\vec{\ell}_T^2}\Bigg\}\left[{\ue}^{-{\ui}(x_L+(1-z)x_M/z)p^+(y^--y_1^-)}+{\ue}^{-{\ui}(x_L+(1-z)x_M/z)p^+y_2^-}\right].
\eea
\end{widetext}
The first two terms on the right-hand side of Eq.~(\ref{eqn:deri2}) represent the contributions from the soft-soft and hard-hard subprocesses, respectively, which remain the same as in Ref.~\cite{Zhang:2004qm}. The third term represents the contributions from their interferences, which give different quark mass dependence as compared to previous calculations in Ref.~\cite{Zhang:2004qm}. It is worth noting that a factor $1/2$ is needed when they are compared with Eq.~(26) in Ref.~\cite{Zhang:2004qm}.

\subsection{Interference from single and triple scatterings}
To complete the calculation, we also need to consider the asymmetric-cut diagrams (left cut and right cut), which represent interferences between single and triple scatterings. All the possible interference diagrams can be found in Ref.~\cite{Wang:2001ifa}. We can obtain the rescattering part $\bar{H}^\mathrm{D}_\mathrm{R(L)}$ of all those asymmetric-cut diagrams in the guidance of the gauge invariance as demonstrated above. The calculation techniques follow exactly the same as those in the double scattering process presented in the previous subsection. Thus we neglect the details and list the final results below,
\begin{widetext}
\bea
\left.\dfrac{\partial^2\bar{H}^\mathrm{D}_\mathrm{L}}{\partial k_{2T}^{\alpha}\partial k_{3T}^{\beta}}\right |_{\substack{k_{2T}=0\\k_{3T}=0}}\!\!&=\!\!\int{\ud}\vec{\ell}_{T}^2\frac{\alpha_s}{2\pi}\frac{1+z^2}{1-z}\frac{2\mathrm{C_A}\vec{\ell}_T^4}{\left[\vec{\ell}_T^2+(1-z)^2M^2\right]^4}{\ue}^{{\ui}(x+x_L+(1-z)x_M/z)p^+y^-}\frac{2\pi\alpha_s}{N_c}\theta(y^--y_1^-)\theta(y_1^--y_2^-)\tilde{H}^\mathrm{D}_\mathrm{L},\label{eqn:aclderi}\\
\left.\dfrac{\partial^2\bar{H}^\mathrm{D}_\mathrm{R}}{\partial k_{2T}^{\alpha}\partial k_{3T}^{\beta}}\right |_{\substack{k_{2T}=0\\k_{3T}=0}}&=\int{\ud}\vec{\ell}_{T}^2\frac{\alpha_s}{2\pi}\frac{1+z^2}{1-z}\frac{2\mathrm{C_A}\vec{\ell}_T^4}{\left[\vec{\ell}_T^2+(1-z)^2M^2\right]^4}{\ue}^{{\ui}(x+x_L+(1-z)x_M/z)p^+y^-}\frac{2\pi\alpha_s}{N_c}\theta(-y_2^-)\theta(y_2^--y_1^-)\tilde{H}^\mathrm{D}_\mathrm{R},
\label{eqn:acrderi}
\eea
where 
\bea
\tilde{H}^\mathrm{D}_\mathrm{L}=&-2\left[\frac{\mathrm{C_F}}{\mathrm{C_A}}(1-z)^2+z\right]\left[(1+z)^2+(1-z)^4\frac{M^2}{\vec{\ell}_T^2}\right]\frac{(1-z)^2}{1+z^2}\frac{M^2}{\vec{\ell}_T^2}\times\left[1-{\ue}^{-{\ui}(x_L+(1-z)x_M/z)p^+(y^--y_1^-)}\right],\label{eq-l}\\
\tilde{H}^\mathrm{D}_\mathrm{R}=&-2\left[\frac{\mathrm{C_F}}{\mathrm{C_A}}(1-z)^2+z\right]\left[(1+z)^2+(1-z)^4\frac{M^2}{\vec{\ell}_T^2}\right]\frac{(1-z)^2}{1+z^2}\frac{M^2}{\vec{\ell}_T^2}\times\left[1-{\ue}^{-{\ui}(x_L+(1-z)x_M/z)p^+y_2^-}\right].\label{eq-r}
\eea

Comparing to the results in Ref.~\cite{Zhang:2004qm}, one can see that the gauge invariant collinear expansion also leads to additional terms,
\bea
\Delta \tilde{H}_\mathrm{L}^\mathrm{D}=&-\left[z-\frac{1}{2}+\mathrm{\frac{C_F}{C_A}}(1-z)^2\right]\left[(1+z)^2+(1-z)^4\frac{M^2}{\vec{\ell}_T^2}\right]\frac{(1-z)^2}{1+z^2}\frac{M^2}{\vec{\ell}_T^2}\nnu
&\times\left[2-{\ue}^{-{\ui}(x_L+(1-z)x_M/z)p^+(y^--y_1^-)}-{\ue}^{-{\ui}(x_L+(1-z)x_M/z)p^+(y^--y_2^-)}\right],\label{eq-diff-l}\\
\Delta \tilde{H}_\mathrm{R}^\mathrm{D}=&-\left[z-\frac{1}{2}+\mathrm{\frac{C_F}{C_A}}(1-z)^2\right]\left[(1+z)^2+(1-z)^4\frac{M^2}{\vec{\ell}_T^2}\right]\frac{(1-z)^2}{1+z^2}\frac{M^2}{\vec{\ell}_T^2}\nnu
&\times\left[2-{\ue}^{-{\ui}(x_L+(1-z)x_M/z)p^+y_2^-}-{\ue}^{-{\ui}(x_L+(1-z)x_M/z)p^+y_1^-}\right].
\label{eq-diff-r}
\eea
Again, the new correction terms vanish in the soft or massless limit.
\end{widetext}

\section{Modified fragmentation function}
\label{sec-MFF}
Substituting the summation of Eqs. (\ref{eqn:ccderi}), (\ref{eqn:aclderi}), and (\ref{eqn:acrderi}) into Eqs. (\ref{factorized}) and (\ref{eqn:tensorw}) and including the gluon fragmentation processes along with virtual corrections, which can be obtained with the help of the unitarity requirement similarly as in Ref.~\cite{Wang:2001ifa}, one can derive the semi-inclusive hadronic tensor from quark-gluon double scattering and interference between single and triple scatterings,
\begin{widetext}
\bea
\frac{{\ud}W_{\mu\nu}^D}{{\ud}z_H}=&\sum_q \int{\ud}xH_{\mu\nu}^{(0)}(x,p,q,M)\int_{z_H}^1\frac{{\ud}z}{z}D_{Q\to H}\left(\frac{z_H}{z}\right)\frac{\alpha_s}{2\pi}\frac{1+z^2}{1-z}\int {\ud}\vec{\ell}_T^2\frac{\mathrm{C_A}\vec{\ell}_T^4}{\left[\vec{\ell}_T^2+(1-z)^2M^2\right]^4}\nnu
&\times \frac{2\pi\alpha_s}{N_c}T_{qg}^{A,Q}(x, x_L,M^2)+(g-\mathrm{fragmentation})+ (\mathrm{virtual}\,\,\mathrm{corrections}),
\eea
where
\bea\label{eq:Tqgam}
T_{qg}^{A,Q}(x, x_L,M^2)\equiv T_{qg}^{A,C}(x, x_L,M^2)+T_{qg}^{A,L}(x, x_L,M^2)+T_{qg}^{A,R}(x, x_L,M^2),
\eea
\bea
T_{qg}^{A,C}(x, x_L, M^2)=&\int\frac{{\ud}y^-}{2\pi}{\ud}y_1^-{\ud}y_2^-\tilde{H}^\mathrm{D}_\mathrm{C}{\ue}^{{\ui}(x+x_L+(1-z)x_M/z)p^+y^-}\nnu
&\times\frac{1}{2}\langle A\mid\bar{\psi}_q(0)\gamma^+{F_\sigma}^+(y_2^-)F^{+\sigma}(y_1^-)\psi_q(y^-)\mid A\rangle\theta(-y_2^-)\theta(y^--y_1^-),\label{eq: TqgC}\\
T_{qg}^{A,L}(x, x_L, M^2)=&\int\frac{{\ud}y^-}{2\pi}{\ud}y_1^-{\ud}y_2^-\tilde{H}^\mathrm{D}_\mathrm{L}{\ue}^{{\ui}(x+x_L+(1-z)x_M/z)p^+y^-}\nnu
&\times\frac{1}{2}\langle A\mid\bar{\psi}_q(0)\gamma^+{F_\sigma}^+(y_2^-)F^{+\sigma}(y_1^-)\psi_q(y^-)\mid A\rangle\theta(y^--y_1^-)\theta(y_1^--y_2^-),\label{eq: TqgL}\\
T_{qg}^{A,R}(x, x_L, M^2)=&\int\frac{{\ud}y^-}{2\pi}{\ud}y_1^-{\ud}y_2^-\tilde{H}^\mathrm{D}_\mathrm{R}{\ue}^{{\ui}(x+x_L+(1-z)x_M/z)p^+y^-}\nnu
&\times\frac{1}{2}\langle A\mid\bar{\psi}_q(0)\gamma^+{F_\sigma}^+(y_2^-)F^{+\sigma}(y_1^-)\psi_q(y^-)\mid A\rangle\theta(-y_2^-)\theta(y_2^--y_1^-)\label{eq: TqgR}.
\eea

In the course of collinear expansion, we have kept $\vec{\ell}_T$ finite when taking the limit $\vec{k}_T\to0$. Consequently, in the soft rescattering, the gluon field in the twist-4 parton matrix elements in part of Eqs. (\ref{eq: TqgC})--(\ref{eq: TqgR}) carries zero momentum. However, in QCD, the gluon distribution function $xf_g(x)$ is not defined at $x = 0$. As argued in Refs.~\cite{Guo:2000nz, Wang:2001ifa}, this issue is owed to the lack of higher order contributions in the collinear expansion. As a remedy, one can resum a subset of the higher-twist terms in the collinear expansion to restore the phase factors in the form as exp(${\ui}x_Tp^+y^−$), where $x_T\equiv \langle \vec{k}_T^2\rangle/2p^+q^−z$ is related to the intrinsic transverse momentum of the initial gluons; namely, the soft gluon fields in the twist-4 matrix elements carry a resulting fractional momentum $x_T$.

Combined with the single scattering contribution, the semi-inclusive tensor can be rewritten in terms of a modified heavy quark fragmentation function $\tilde{D}_{Q\to H}(z_H,\mu^2)$,
\bea
\frac{{\ud}W_{\mu\nu}}{{\ud}z_H}=\sum_q \int{\ud}x\tilde{f}_q^A(x,\mu_I^2)H_{\mu\nu}^{(0)}(x,p,q,M)\tilde{D}_{Q\to H}(z_H,\mu^2) + \dots
\eea
\end{widetext}
where $\tilde{f}_q^A(x,\mu^2_I)$ is the nuclear quark distribution function that, in principle, should also include the higher-twist contribution from the initial state scattering. In this study, we focus on the effect of final state multiple scattering and neglect the initial state multiple scattering, and thus $\tilde{f}_q^A(x,\mu^2_I)\approx f_q^A(x,\mu^2_I)$ with being $f_q^A(x,\mu^2_I)$ the standard leading twist nuclear quark distribution function. The modified effective heavy quark fragmentation function is defined as
\begin{widetext}
\bea
\tilde{D}_{Q\to H}(z_H,\mu^2)\equiv& D_{Q\to H}(z_H,\mu^2)+\int_0^{\mu^2}\frac{{\ud}\vec{\ell}_{T}^2}{\vec{\ell}_{T}^2+(1-z)^2M^2}\frac{\alpha_s}{2\pi}\int_{z_H}^1
\frac{{\ud}z}{z}\Delta\gamma_{Q\to Qg}(z,x,x_L,\vec{\ell}_{T}^2,M^2)D_{Q\to H}(z_H/z,\mu^2)\nnu
&+\int_0^{\mu^2}\frac{{\ud}\vec{\ell}_{T}^2}{\vec{\ell}_{T}^2+z^2M^2}\frac{\alpha_s}{2\pi}\int_{z_H}^1
\frac{{\ud}z}{z}\Delta\gamma_{Q\to gQ}(z,x,x_L,\vec{\ell}_{T}^2,M^2)D_{g\to H}(z_H/z,\mu^2),
\eea
where $D_{Q\to H}(z_H, \mu^2)$ and $D_{g\to H}(z_H, \mu^2)$ are the heavy quark and gluon fragmentation functions at leading twist, respectively. The modified splitting functions are defined as
\bea
\Delta\gamma_{Q\to Qg}(z,x,x_L,\vec{\ell}_{T}^2,M^2)=&\left[\frac{1+z^2}{(1-z)_+}T_{qg}^{A,Q}(x, x_L,M^2)+\delta(1-z)\Delta T_{qg}^{A,Q}(x, \vec{\ell}_{T}^2,M^2)\right]\nnu
&\times\frac{2\pi\alpha_s\mathrm{C_A}\vec{\ell}_{T}^4}{\left[\vec{\ell}_{T}^2+(1-z)^2M^2\right]^3N_c f_q^A(x,\mu_I^2)},
\eea
\bea
\Delta\gamma_{Q\to gQ}(z,x,x_L,\vec{\ell}_{T}^2,M^2)=\Delta\gamma_{Q\to Qg}(1-z,x,x_L,\vec{\ell}_{T}^2,M^2),
\eea
where
\bea
\Delta T_{qg}^{A,Q}(x, \vec{\ell}_{T}^2,M^2)\equiv\int_0^1{\ud}z\frac{1}{1-z}\left[2T_{qg}^{A,Q}(x, x_L,M^2)|_{z=1}-(1+z^2)T_{qg}^{A,Q}(x, x_L,M^2)\right].
\eea
\end{widetext}

\section{Heavy quark energy loss}
\label{sec-energy loss}
As shown in Eqs.~(\ref{eq-diff-sh}), (\ref{eq-diff-hs}), (\ref{eq-diff-l}), and (\ref{eq-diff-r}), the new correction terms in this calculation are all proportional to $(1-z)^2M^2$, which vanish in the soft or massless limit. Therefore, for light quark energy loss calculation as shown in Refs.~\cite{Guo:2000nz,Wang:2001ifa,Zhang:2003yn}, the final result is gauge invariant, and the phenomenological applications based on this result remain the same. The heavy quark energy loss as calculated in Refs.~\cite{Zhang:2003wk,Zhang:2004qm} is complete and gauge invariant only in the soft gluon radiation limit, which has been employed in phenomenological study of heavy meson production in heavy-ion collisions \cite{cao2013heavy,PhysRevC.92.024907,cao2018heavy}. For a more complete phenomenological investigation of heavy quark energy loss beyond soft gluon limit, one should consider the issue of gauge invariance and the results obtained in this study should be used instead.  

In order to quantitatively estimate the new correction terms in heavy quark energy loss, we consider the leading contribution in the limit of a large nuclear size, which is proportional to $A^{2/3}$ due to non-Abelian Laudau-Pomeranchuk-Migdal (LPM) interference in the twist-4 contributions. Following the same ansatz as in Refs.~\cite{Zhang:2003wk,Zhang:2004qm} for the nonperturbative twist-4 parton matrix element, we assume a factorized form in the limit $x_L\ll x$,
\begin{equation}
\begin{aligned}
\!\!\!\!T_{qg}^{A,C}(x, x_L, M^2)\approx\frac{\tilde{\mathrm{C}}}{x_A}f_q^A(x)(1-{\ue}^{-\tilde{x}_{L}^2/x_A^2})a(z,M^2/\vec{\ell}^2_T),
\label{eq-Tqg}
\end{aligned}
\end{equation}
where $\tilde{x}_L=x_L+zx_M/(1-z)$, $x_A\equiv1/m_NR_A$, and the coefficient $\mathrm{\tilde{C}}$ is proportional to the gluon distribution inside a nucleon, whose value can be taken as $\mathrm{\tilde{C}} = 0.0060$ as determined by fitting to data on light hadron production in DIS off nuclear targets according to Ref.~\cite{Wang:2002ri}. The suppression factor $1-{\ue}^{-\tilde{x}_L^2/x_A^2}$ due to the LPM interference arises from the phase factor in Eq.~(\ref{eqn:deri2}) integrated with a Gaussian nuclear distribution of a radius $R_A$. In Eq.~(\ref{eq-Tqg}), 
\begin{widetext}
\begin{equation}
\begin{aligned}
a(z,M^2/\vec{\ell}^2_T)=&\frac{(1+z)}{2}-\frac{2(1-z)^3z(1+z)}{1+z^2}\frac{M^2}{\vec{\ell}_T^2}+\frac{(1-z)^4(3z^3-5z^2+7z-1)}{2(1+z^2)}\frac{M^4}{\vec{\ell}_T^4}\\
&+\mathrm{\frac{2C_F}{C_A}}\left[(1+z)^2+(1-z)^4\frac{M^2}{\vec{\ell}_T^2}\right]\frac{(1-z)^4}{1+z^2}\frac{M^2}{\vec{\ell}_T^2},
\label{eq-a1}
\end{aligned}
\end{equation}
which differs from Ref.~\cite{Zhang:2004qm} by
\bea
\Delta a\equiv a-a^{\text{Ref.~\cite{Zhang:2004qm}}}
=\left [z-\frac{1}{2}+\mathrm{\frac{C_F}{C_A}}(1-z)^2\right ]\left [(1+z)^2+(1-z)^4\frac{M^2}{\vec{\ell}_T^2}\right ]\frac{(1-z)^2}{1+z^2}\frac{M^2}{\vec{\ell}_T^2}.
\label{eq-diff-a}
\eea
In massless limit $M=0$, $a$ reduces to $(1+z)/2$ and reproduces the result in Ref.~\cite{Zhang:2003yn}. 

With the parametrized form of the twist-4 matrix element as in Eq.~(\ref{eq-Tqg}), one can then estimate the averaged heavy quark energy loss, which is defined as the fractional energy carried by the radiated gluon,
\begin{equation}
\begin{aligned}
\langle\Delta z_g^Q\rangle(x_B, \mu^2)=&\int_0^{\mu^2}{\ud}\vec{\ell}_{T}^2\int_0^1{\ud}z \frac{\alpha_s}{2\pi}(1-z)\frac{\Delta\gamma_{Q\to Qg}(z,x_B,x_L,\vec{\ell}_{T}^2)}{\vec{\ell}_{T}^2+(1-z)^2M^2}\\
=&\frac{\mathrm{\tilde{C}C_A\alpha_s^2}x_B}{N_cQ^2x_A}\int_0^1{\ud}z\frac{1+(1-z)^2}{z(1-z)}\int_{\tilde{x}_M}^{\tilde{x}_\mu}{\ud}\tilde{x}_L\frac{(\tilde{x}_L-\tilde{x}_M)^2}{\tilde{x}^4_L}(1-{\ue}^{-\tilde{x}_{L}^2/x_A^2})a(z,M^2/\vec{\ell}^2_T),
\label{eq-deltaz}
\end{aligned}
\end{equation}
\end{widetext}
where $\tilde{x}_M=zx_M/(1-z)$ and $\tilde{x}_\mu=\mu^2/2p^+q^-z(1-z)+ \tilde{x}_M=x_B/z(1-z)+ \tilde{x}_M$ if we choose the factorization scale $\mu^2 = Q^2$, and the constraint of $\tilde{x}_L$ integration comes from the requirement of $x_L<1$ as in Ref.~\cite{Zhang:2004qm}.  

\begin{figure*} 
\centering
\subfloat[][\empty]{
\begin{minipage}[t]{0.4\linewidth} 
\centering 
\includegraphics[width=2.8in]{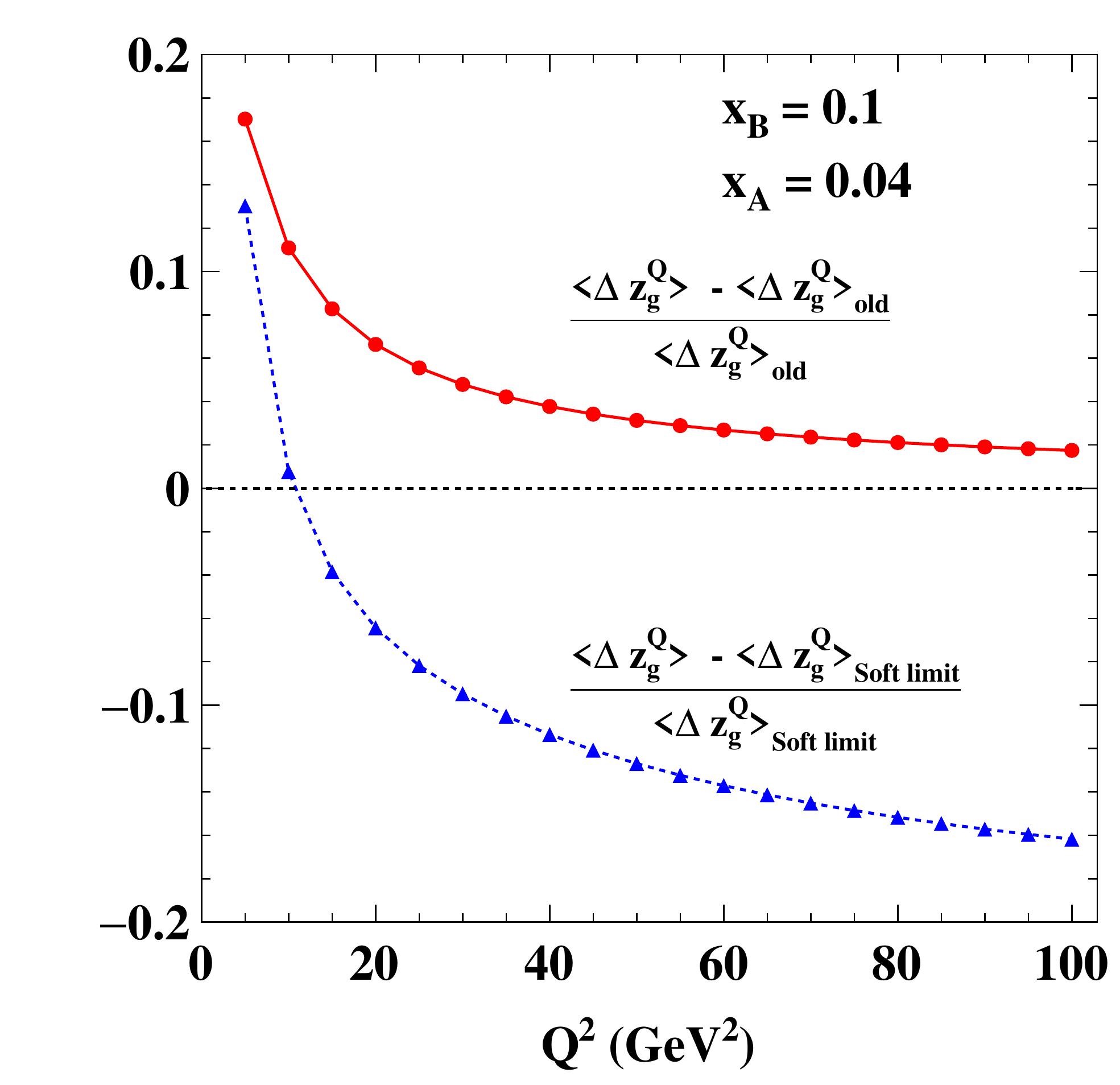} 
\end{minipage}%
} 
\subfloat[][\empty]{
\begin{minipage}[t]{0.4\linewidth} 
\centering 
\includegraphics[width=2.8in]{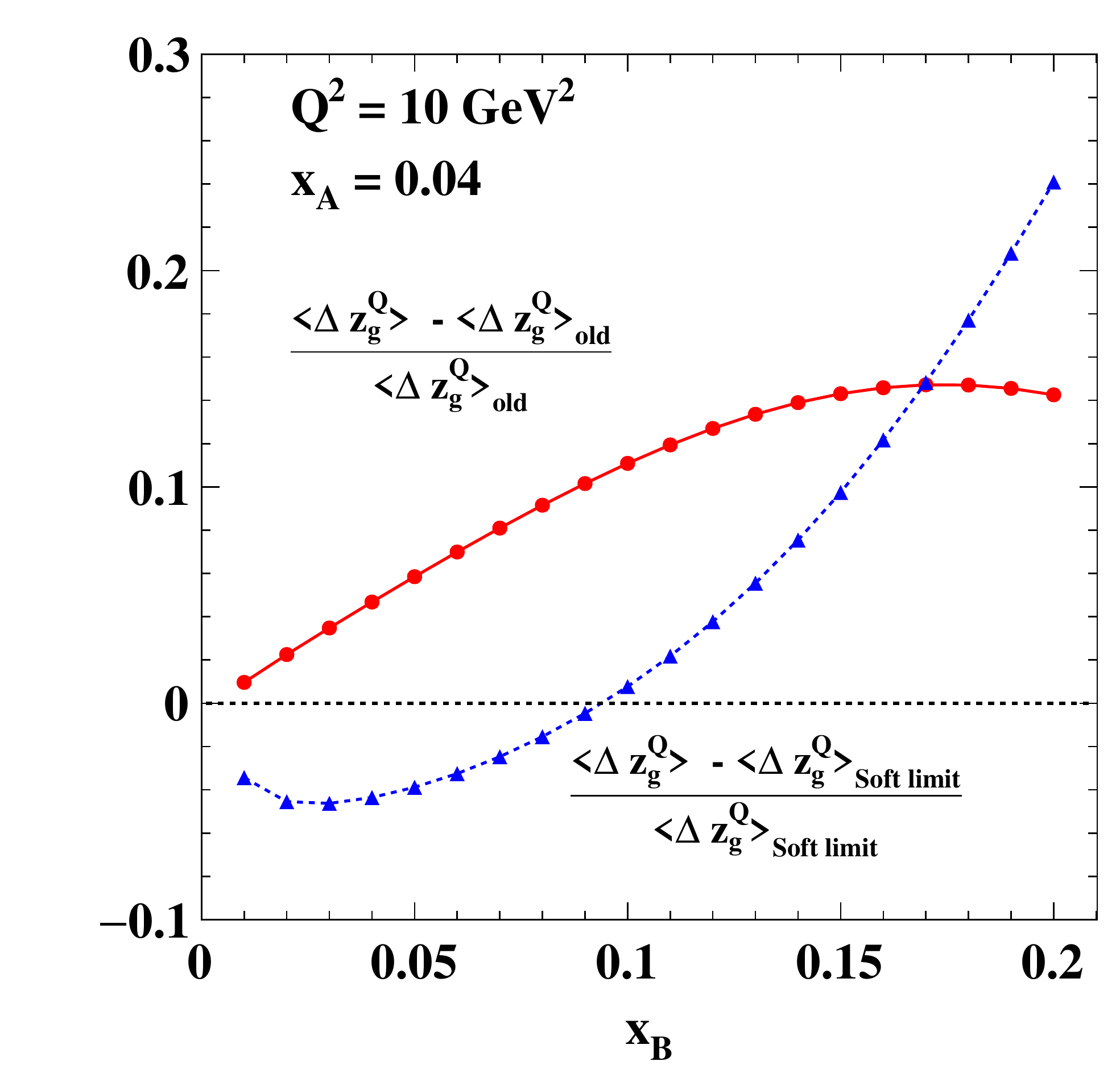} 
\end{minipage} 
}\caption{ The $Q^2$ and $x_B$ dependence of the relative correction of charm quark energy loss $\delta\langle\Delta z_g^Q\rangle/\langle\Delta z_g^Q\rangle$ as compared to that from old work~\cite{Zhang:2004qm} (red solid) and with soft gluon approximation (blue dashed).}
\label{fig-DELTAE}
\end{figure*}

\begin{figure*} 
\centering
\subfloat[][\empty]{
\begin{minipage}[t]{0.4\linewidth} 
\centering 
\includegraphics[width=2.8in]{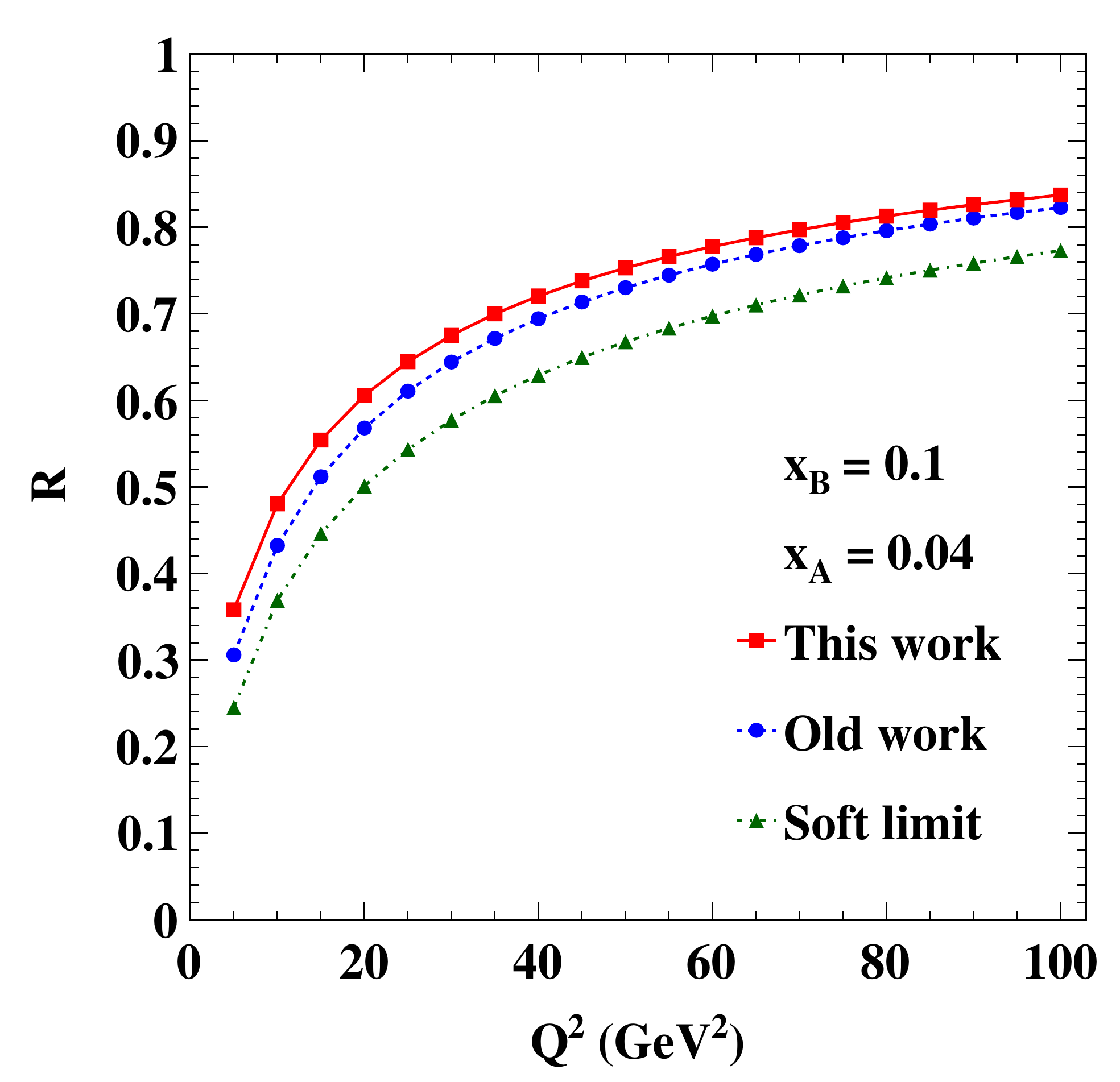} 
\end{minipage}%
} 
\subfloat[][\empty]{
\begin{minipage}[t]{0.4\linewidth} 
\centering 
\includegraphics[width=2.8in]{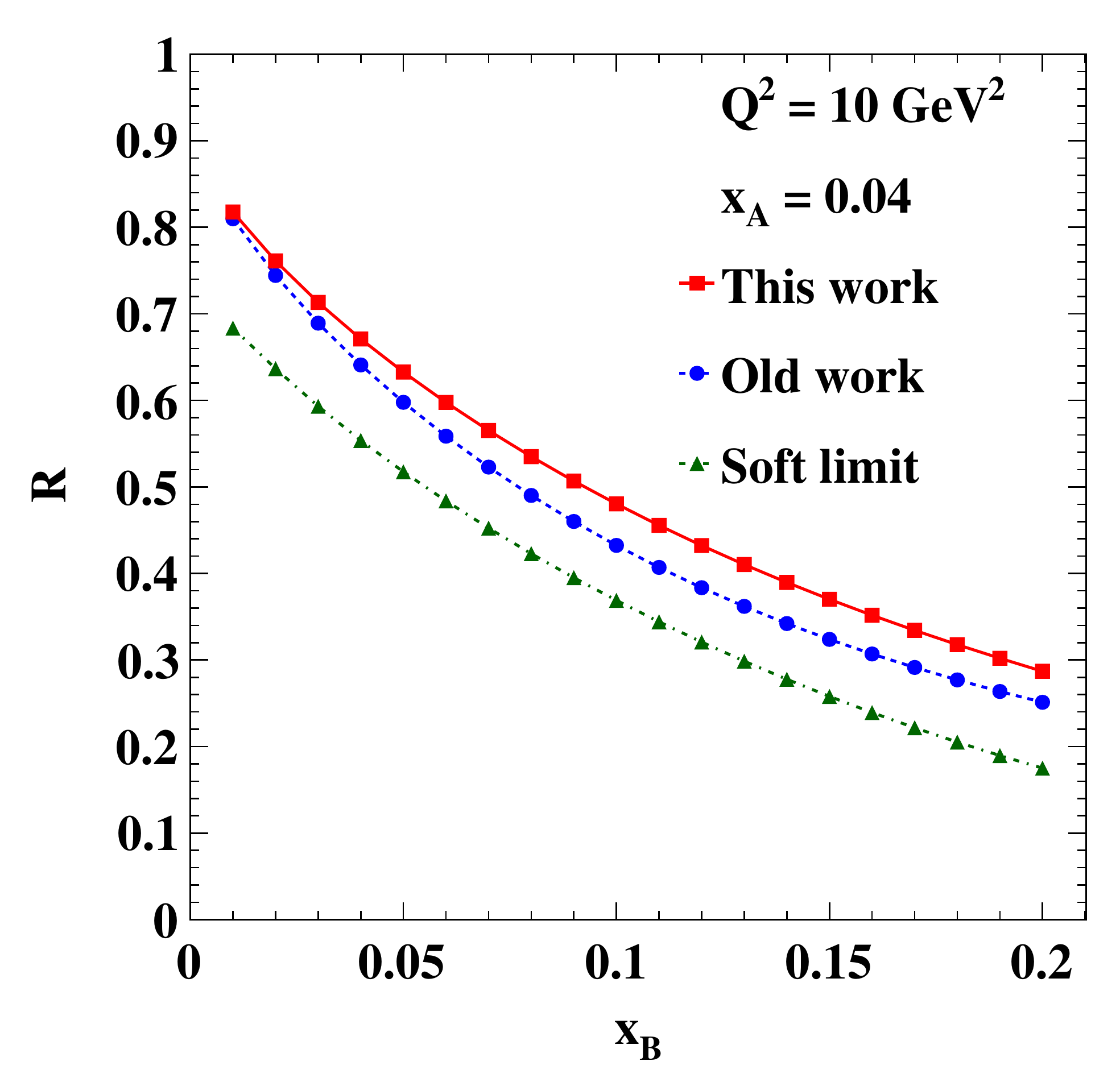} 
\end{minipage} 
}\caption{ The $Q^2$ and $x_B$ dependence of the ratio between charm quark and light quark energy loss $R$ from this work (red solid), old work~\cite{Zhang:2004qm} (blue dashed), and soft limit (green dashed-dotted).}
\label{fig-R}
\end{figure*}

To evaluate heavy quark energy loss numerically, we choose charm quark mass $M = 1.5$ GeV and $x_A = 0.04$ for a nucleus with a radius $R_A=5$ fm. In the left panel of Fig.~\ref{fig-DELTAE}, we show in the red solid curve the relative difference between the new result in this study and that in Ref.~\cite{Zhang:2004qm} as a function of $Q^2$ with fixed $x_B=0.1$. One can see that the new correction term due to consideration of gauge invariance leads to significant additional contribution to the heavy quark energy loss in the small $Q^2$ region. However, the difference becomes negligible in the large $Q^2$ region. This is understandable from Eq.~(\ref{eq-diff-a}). The difference is proportional to $M^2$, and therefore suppressed in the large $Q^2$ region like any other higher-twist effect. The difference between our new result and the commonly used soft gluon radiation limit is shown in the blue dashed curve, which is as large as $16\%$ in the large $Q^2$ region. It is therefore important to consider contributions beyond the soft limit in more precise phenomenological studies of heavy quark energy loss in heavy-ion collisions. Shown in the right panel of Fig.~\ref{fig-DELTAE} is the relative difference between our new results and the previous calculation in Ref.~\cite{Zhang:2004qm} as a function of $x_B$ for fixed $Q^2= 10$ GeV$^2$. The difference becomes significant for large values of  $x_B$ (small initial heavy quark energy) as shown in the red solid curve. The contribution beyond the soft gluon limit is also appreciable at large $x_B$ (small heavy quark energy) as shown in the blue dashed curve.

In order to discuss the difference between heavy and light quark energy loss, we show the ratio of charm quark and light quark energy loss $R=\langle\Delta z_g^Q\rangle(x_B,\mu^2)/\langle\Delta z_g^q\rangle(x_B,\mu^2)$ in Fig.~\ref{fig-R}, where the light quark energy loss $\langle\Delta z_g^q\rangle(x_B,\mu^2)$ can be obtained by setting $M=0$ in Eq. (\ref{eq-deltaz}). In Fig.~\ref{fig-R}, we show the dependence of $R$ with $Q^2$ for fixed $x_B=0.1$ (left panel)
and with $x_B$ for fixed $Q^2=10$ GeV$^2$ (right panel). We observe the reduction of heavy quark energy loss due to the effect of the dead cone in our new result (red solid curve) as compared to that in Ref.~\cite{Zhang:2004qm} (blue dashed curve).  Such a reduction is due to the heavy quark mass and therefore should disappear at high $Q^2$ and large initial quark energy (small $x_B$). Again, contributions from beyond the soft gluon limit are shown to be important by comparing the red solid curve (our new result) and the green dashed-dotted curve [soft gluon radiation limit, which is also needed to be taken in light quark energy loss $\langle\Delta z_g^q\rangle(x_B,\mu^2)$ at the same time].

\section{Summary}
\label{sec-sum}

In this paper, we revisited a series of studies~\cite{Guo:2000nz,Wang:2001ifa,Zhang:2003yn,Zhang:2003wk,Zhang:2004qm} on quark energy loss induced by multiple parton scattering in DIS off a nuclear target, using the recently improved framework of the generalized factorization formalism for twist-4 processes. By performing the gauge invariant collinear expansion, we found that the light quark energy loss is not affected, but new correction terms arise for heavy quark energy loss beyond the soft gluon limit. The correction terms come from the interference between soft and hard rescatterings. In the soft gluon limit, the new correction terms can be safely neglected since they are proportional to $(1-z)^2$. This validates the phenomenological implementations of heavy quark energy loss from high-twist calculations in heavy-ion collisions as have been presented in Refs.~\cite{cao2013heavy,PhysRevC.92.024907,cao2018heavy}.

To demonstrate the significance of the correction terms, we evaluated numerically the heavy quark energy loss and compared with that in Ref.~\cite{Zhang:2004qm}. We found significant correction in the small $Q^2$ and large $x_B$ (small heavy quark energy) regions. Our new result was also compared with that with soft gluon approximation. The noticeable difference between these two suggests the importance of implementing the complete result (beyond soft limit) for more precise calculation of heavy flavor jet quenching in heavy-ion collisions. This also has phenomenological consequences in precise extraction of the jet transport coefficient from light and heavy flavor data in heavy-ion collisions.

\acknowledgments
This work is supported by the Helmholtz Graduate School HIRe for FAIR (Y. D.), by the F\&E Programme of GSI Helmholtz Zentrum f$\ddot{\mathrm{u}}$r Schwerionenforschung GmbH, Darmstadt (Y. D.), by the Giersch Science Center (Y. D.), by the Walter Greiner Gesellschaft zur F$\ddot{\mathrm{o}}$rderung der physikalischen Grundlagenforschung e.V., Frankfurt (Y. D.), and by the AI grant of SAMSON AG, Frankfurt (Y. D.); by National Natural Science Foundation of China under Grants No. 11475085, No. 11535005, No. 11690030 (Y. D. and H. Z.), and No. 11221504 (Y. H. and X.-N.W.), and National Major state Basic Research and Development of China under Grants No. 2016YFE0129300 (Y. D. and H. Z.) and No. 2014CB845404 (Y. H. and X.-N.W.); and the U.S. Department of Energy under Awards No. DE-FG02-91ER40684,  No. DE-AC02-06CH11357 (H. X.), and No. DE-AC02-05CH11231 (X.-N.W.), and the U.S. National Science Foundation (NSF) under Grant No. ACI-1550228 (JETSCAPE; X.-N.W.). H. X. is grateful for the hospitality of the Institute of Particle Physics at Central China Normal University during the completion of this manuscript.

\bibliographystyle{apsrev4-1}
\bibliography{biblio}

\end{document}